\documentclass{PoS}
\usepackage{amsmath}

\title{Exact Results in AdS$_4$/CFT$_3$}

\ShortTitle{Exact Results in AdS$_4$/CFT$_3$}

\author{\speaker{Silvia Penati}\thanks{This talk is based on work done in collaboration with M.S. Bianchi, L. Griguolo, M. Leoni, A. Mauri, M. Preti, D. Seminara}\\
       Universit\`a degli studi di Milano Bicocca and INFN, Sezione di Milano - Bicocca, \\ Piazza
della Scienza 3, 20161, Milano, Italy\\
        E-mail: \email{silvia.penati@mib.infn.it}}

\abstract{I review recent results concerning the construction of generalized ``latitude'' Wilson loops in ABJM theory. In particular, the parametric Matrix Model determining these operators exactly is presented, as well as the exact prescription for computing different types of Bremsstrahlung functions from circular Wilson loops. In this context the physical meaning of framing in non-topological three-dimensional theories is clarified.}

\FullConference{Corfu Summer Institute 2019 "School and Workshops on Elementary Particle Physics and Gravity" (CORFU2019)\\
		31 August - 25 September 2019\\
		Corfù, Greece}

\newcommand{\beq}{\begin{equation}}
\newcommand{\bea}{\begin{eqnarray}}
\newcommand{\eea}{\end{eqnarray}}
\newcommand{\eeq}{\end{equation}}

\newcommand{\G}{\Gamma}

\def \g {\gamma}

\def \G {\Gamma}

\def \e {\epsilon}

\def \m {\mu}
\def \n {\nu}

\def \l {\lambda}
\def \L {\Lambda}

\def \th {\theta}

\def \Tr {{\textrm{Tr}}}

\begin{document}

\section{Introduction}

In superconformal gauge theories BPS Wilson loops (WLs) can be defined, which are non-local, gauge covariant or invariant Wilson-type operators that preserve a fraction of the superconformal charges. These operators are in general non-protected against quantum corrections and play a ubiquitous role in testing the AdS/CFT correspondence. In fact, their vacuum expectation values (vevs) undergo a non-trivial flow between weak and strong coupling regimes and when localization techniques  are available for their exact determination they provide exact interpolating functions. 
In supeconformal field theories (SCFTs) BPS WLs are also related to other physical observables, like the cusp anomalous dimension and the Bremsstrahlung function, which can be determined in terms of the WL expectation value. Since, alternatively, these quantities can in principle be computed by using integrability techniques, the study of BPS WLs can also be instrumental for testing integrability underlying the AdS/CFT correspondence. 

From a different perspective, BPS WLs can be interpreted as dynamical one-dimensional defects embedded in higher dimensional theories. In particular, interest has recently grown in studying the structure of these lower dimensional superconformal defects through the evaluation of correlation functions of local operators inserted on the Wilson contour.

BPS Wilson operators have been introduced and studied first in the prototypical example of four-dimensional ${\cal N}=4$ $SU(N)$ SYM theory. For the generalized circular 1/2 BPS Wilson-Maldacena operator \cite{Maldacena:1998im,Rey:1998ik} which includes couplings to scalar fields, a gaussian Matrix Model on the $S^4$ sphere has been found \cite{Erickson:2000af,Drukker:2000rr,Pestun:2007rz}, which computes exactly its expectation value $\langle W \rangle$. This function interpolates between the perturbative result at weak coupling \cite{Erickson:2000af} and the strong coupling prediction provided by a dual string configuration in AdS$_5 \times $S$^5$ \cite{Maldacena:1998im,Drukker:2000rr,Drukker:1999zq}. An exact prescription has been then proposed for computing the Bremsstrahlung function $B$, the physical observable measuring the energy lost by a massive quark slowly moving in the gauge background, in terms of $\langle W \rangle$ \cite{Correa:2012at,Fiol:2012sg}. This function also enters the small angle expansion of the cusp anomalous dimension $\G_{cusp}(\phi)\sim - B\phi^2$, which in turn controls the short distance divergences of a WL in the proximity of a cusp featured by an angle $\phi$, according to the universal behaviour $\langle W(\phi) \rangle \sim \exp{(-\G_{cusp}}(\phi) \log{\tfrac{\L}{\e})}$ (here $\L$ and $\e$ are IR and UV regulators, respectively). Exploiting integrability, the same quantities have been determined by solving a system of TBA equations \cite{Bombardelli:2009ns,Gromov:2009bc,Arutyunov:2009ur} with boundaries and in the near-BPS limit, for a generalized cusp with the insertion of R-charged chiral operators on the tip of the cusp \cite{Drukker:2012de,Correa:2012hh,Gromov:2012eu,Gromov:2013qga}. 
More general results away from the BPS point has also been obtained by the use of the quantum spectral curve techniques \cite{Bajnok:2013sya,Gromov:2015dfa}. Large families of less BPS WLs in ${\cal N}=4$ SYM theory have been also introduced \cite{Drukker:2006ga,Drukker:2007dw,Drukker:2007yx,Drukker:2007qr}, which depend on constant parameters featuring the internal coupling with the scalars and/or the contour, and interpolate between WLs with different degree of supersymmetry. They have been computed at weak and strong coupling and exactly via localization \cite{Giombi:2009ms,Pestun:2009nn,Giombi:2009ds}. 

More generally, in four dimensions this approach has been developed for studying BPS WLs in ${\cal N} \geq 2$ SYM theories. The exact  result for $\langle W \rangle$ is still provided by a Matrix Model on $S^4$ \cite{Pestun:2007rz}, which includes a non-trivial one-loop determinant and an instanton factor and is no longer gaussian.  Remarkably, a prescription for computing the $B$ function  has been given in \cite{Fiol:2015spa} in terms  of circular WLs on the squashed sphere \cite{Hama:2012bg}. This prescription has been tested up to three loops in \cite{Gomez:2018usu}, and it has been recently proved in general by exploiting algebraic properties of correlation functions induced by the residual superconformal invariance on the Wilson line \cite{Bianchi:2018zpb,Bianchi:2019dlw}.  

This kind of investigation has been extended to three-dimensional models. In this proceedings I will review recent results regarding BPS Wilson loops and related observables in super-Chern-Simons-matter theories, with particular emphasis on the ${\cal N}=6$, $U(N) \times U(M)$ ABJ(M) models \cite{ABJM,ABJ}. The discussion is primarily based on papers \cite{Bianchi:2014laa,Bianchi:2017svd,Bianchi:2018bke}. For a broader collection of results and an exhaustive list of references on WLs in Chern-Simons-matter theories we refer to the recent review \cite{Drukker:2019bev}. 

In three dimensions the spectrum of BPS WLs is much richer than in four dimensions. In fact, due to simple dimensional reasons, not only scalar but also fermion matter can be used to build up generalized BPS loop operators. There are in fact two prototypes of supersymmetric WLs: One (the bosonic WL) is associated to a generalized gauge connection that includes couplings to quadratic terms in the bosonic fields, and preserves at most 1/6 of the original supersymmetries \cite{Berenstein:2008dc,Drukker:2008zx,Chen:2008bp}. It should be dual to fundamental strings smeared along a CP$^1$ inside CP$^3$. The second one (the fermionic WL) is featured by a gauge superconnection which includes also couplings to fermions \cite{Drukker:2009hy}. The inclusion of fermions promotes the operator to be at most 1/2 BPS, which is dual to a fundamental string on AdS$_4 \times$ CP$^3$. 

Though the two types of operators preserve different portions of supersymmetry, the fermionic WL is cohomologically equivalent to a linear combination of the bosonic ones. It then follows that at quantum level they are indistinguishable, and a Matrix Model obtained by localizing with their cohomological charge computes both of them. Indeed, this Matrix Model has been proposed in \cite{Kapustin:2009kz} together with its weak coupling expansion, whereas an exact expression at large $N$ in the strong coupling limit has been found using topological strings \cite{Marino:2009jd,Drukker:2010nc,Drukker:2011zy} and the Fermi gas approach \cite{Marino:2011eh,Klemm:2012ii}. 
 
Having different types of generalized WLs allows to construct different non-BPS observables starting from them. Generalized cusps formed with 1/6 BPS  or 1/2 BPS rays are actually different \cite{Griguolo:2012iq,Lewkowycz:2013laa} and, consequently, different Bremsstrahlung functions can be defined and potentially evaluated exactly, as we will review. 

The rest of the paper is structured as follows. In section \ref{WL} we briefly recall the definition of BPS WLs in four and three dimensions. In section 
\ref{latitude} these definitions are generalized to define two one-parameter classes of BPS operators, the bosonic and the fermionic ``latitudes''. We evaluate them perturbatively, discuss their cohomological equivalence, the concept of framing in non-topological theories and their framing dependence. For these parametric WLs, in section \ref{MM} we propose a Matrix Model to compute them exactly. This is a parametric Matrix Model that should arise from localizing the path integral with the parameter-dependent supercharge that enters the cohomological equivalence between fermionic and bosonic WLs. We compute the Matrix Model at large $N$ in the strong coupling regime by applying Fermi gas techniques and discuss the consistency of our proposal. Section \ref{B} is devoted to the definition of the different Bremsstrahlung functions associated to bosonic and fermionic WLs and contains the exact prescription to compute the $B$'s in terms of latitude WLs. Remarkably, as discussed in section \ref{Bframing}, the fermionic Bremsstrahlung function turns out to be entirely determined by the framing function of the bosonic WL. This gives a new physical meaning to regularization-dependent framing factors in the case of non-topological theories. Finally, we discuss open questions and future perspectives in section \ref{conclusions}.

\section{BPS Wilson loops}\label{WL}

In supersymmetric gauge theories, ordinary Wilson loops 
\beq\label{W}
W = \Tr \, P \, e^{-i \int_\G dx^\mu A_\mu}
\eeq
break all supersymmetries, since there is no choice of the contour $\G$ that renders the holonomy of the gauge connection supersymmetry invariant\footnote{A manifestly supersymmetric version of (\ref{W}) can be formulated in superspace, in terms of the integral of a superconnection on a supercontour \cite{Gates:1976rk}. A rheonomic formulation of these operators has been recently proposed in \cite{Cremonini:2020mrk}.}. However, if the spectrum of the theory contains matter fields in the adjoint representation of the gauge group, the connection can be generalised to include couplings to these extra fields. These internal couplings and the contour can then be suitably triggered in order to make the operator (locally) preserving a fraction of the supersymmetry charges. 

The prototypical example is the generalized Wilson-Maldacena operator introduced in four-dimensional ${\cal N}=4$, $SU(N)$ SYM  \cite{Maldacena:1998im,Rey:1998ik}. In Euclidean signature and in fundamental representation it is defined as
\beq\label{WL4D}
W = \Tr \, P \, e^{-i \int_\Gamma d\tau {\cal L}(\tau)}
\; ,  \qquad  {\cal L}(\tau) = \dot{x}^\mu A_\mu + i |\dot{x}| \theta_I (\tau) \Phi^I   (\tau) 
\eeq
where $\theta_I(\tau)$, $I=1, \cdots , 6$, drive the (local) coupling to the six scalar fields $\Phi^I$ of the theory and $\Tr$ means the trace taken in fundamental representation of the gauge group. This expression can be easily obtained by dimensional reduction of an ordinary WL in ten dimensions or, alternatively, arises in a spontaneous $SU(N+1) \to SU(N) \times U(1)$ symmetry breaking mechanisms driven by the non-vanishing vev of some scalars and describes the phase associated to the dynamics of a massive W-boson moving in the gauge background \cite{Maldacena:1998im,Rey:1998ik,Drukker:1999zq}. 

When the contour $\Gamma$ is a closed loop this operator is gauge invariant. For a suitable choice of $\Gamma$ and $\theta_I$ (satisfying $\theta^2 = 1$) it can be shown to preserve a fraction of supercharges. In particular, when $\theta_I$ is constant and $\Gamma$ is the maximal circle in $S^2 \subset S^4$ this is 1/2 BPS, 
i.e. it preserves half of the superconformal charges. 
According to the AdS/CFT correspondence, this operator has a dual description in terms of fundamental strings ending on the $\Gamma$ contour at the boundary of AdS$_5$.  
 
Operators (\ref{WL4D}) are in general non--protected and their expectation value  
\beq
\langle W \rangle \sim \int D[A, \hat{A}, C, \bar{C}, \psi, \bar{\psi}] \; e^{-S} \; \,  \Tr \, P \, e^{ - i \int_\Gamma d\tau {\cal L}(\tau)  }  
\eeq
depends non-trivially on the coupling constant $\l = g^2 N$ of the theory. For circular contours, they can be computed at weak couplings by ordinary perturbation theory \cite{Erickson:2000af,Drukker:2000rr}, whereas at strong couplings one can use holographic methods \cite{Drukker:1999zq}. According to this prescription the vev is given by $\langle W \rangle = Z_{\rm string}$ where $Z_{\rm string}$ is the string partition function evaluated at the minimal area worldsheet ending on the WL contour. In addition, for theories with ${\cal N} \geq 2$ supersymmetry, $\langle W \rangle$ can be computed at any finite value of the coupling using localization techniques \cite{Drukker:2000rr,Pestun:2007rz,Giombi:2009ms,Pestun:2009nn,Giombi:2009ds}. In this approach the path integral defining the vev localizes to a Matrix Model which can be solved exactly. For the 1/2 BPS WL in${\cal N}=4$ SYM theory, at finite $N$ one obtains
\beq\label{4Dexact}
\langle W_{1/2} \rangle = \frac{1}{N} \, L^1_{N-1} \! \! \left( - \frac{\l}{4N} \right) \, e^{\frac{\l}{8N}}
\eeq
where $L^1_{N-1}$ is the modified Laguerre polynomial. In the large $N$ limit this result coincides with the resumation of the perturbative series of ladder diagrams \cite{Erickson:2000af,Drukker:2000rr}. Moreover, its leading order expansion at strong coupling coincides with the string theory prediction \cite{Maldacena:1998im,Drukker:1999zq,Drukker:2005cu}, as well as the one-loop correction \cite{Drukker:2000ep,Kruczenski:2008zk,Buchbinder:2014nia}. Therefore, the Matrix Model result (\ref{4Dexact}) provides an exact interpolating function that can be used to check the AdS/CFT correspondence.

\section{``Latitude'' bosonic and fermionic BPS Wilson loops  in ABJM theory}\label{latitude}

We now introduce the main subject of this review, that is WLs in three-dimensional Chern-Simons-matter theories. We will primarily focus on the ${\cal N}=6$ ABJM model \cite{ABJM}, though most of the discussion that follows has a simple generalisation (with some {\em distinguo}) to the more general ABJ theory \cite{ABJ}.  

The ${\cal N}=6$ ABJM theory is a three-dimensional $U(N)_k \times U(N)_{-k}$  Chern-Simons-matter theory\footnote{The subscript $k$ indicates the Chern-Simons level. It is  an integer, as required by gauge invariance.}, whose field content is given by $A_\mu$, 
$\hat{A}_\mu$ gauge vectors minimally coupled to $SU(4)$ complex scalars $C_I$, $\bar{C}^I$ and corresponding fermions $\bar{\psi}^I$, $\psi_I$, $I=1, \dots, 4$
 in the (anti)bifundamental representation of the gauge group, and subject to a non-trivial potential. The total action reads
\beq
 S = S_{\mathrm{CS}} + S_{\mathrm{mat}} + S_{\mathrm{pot}}^{\mathrm{bos}} + S_{\mathrm{pot}}^{\mathrm{ferm}} 
\eeq
where
\begin{eqnarray}   
&& \hspace{-1.4cm}   S_{\mathrm{CS}} =\frac{k}{4\pi i}\int d^3x\,\varepsilon^{\mu\nu\rho} \Big\{  {\rm Tr}  \left( A_\mu\partial_\nu A_\rho+\frac{2}{3} i A_\mu A_\nu A_\rho\! \right) \!-\! {\rm Tr}  \!\left(\hat{A}_\mu\partial_\nu \hat{A}_\rho+\frac{2}{3} i \hat{A}_\mu \hat{A}_\nu \hat{A}_\rho \right) \Big\}\nonumber \\
&&~~\nonumber \\
&& \qquad S_{\mathrm{mat}} = \int d^3x \, {\rm Tr} \Big[ D_\mu C_I D^\mu \bar{C}^I - i \bar{\Psi}^I \g^\mu D_\mu \Psi_I \Big] 
\nonumber
\end{eqnarray}
Here we have defined covariant derivatives 
\beq
D_\mu C_I = \partial_\mu C_I + i A_\mu C_I - i C_I \hat{A}_\mu \qquad , \qquad D_\mu \bar{C}^I = \partial_\mu \bar{C}^I - i \bar{C}^I  A_\mu + i \hat{A}_\mu \bar{C}^I
\eeq
and similarly for fermions. We avoid writing explicitly the complicated expressions of the potential terms. $S_{\mathrm{pot}}^{\mathrm{bos}}$ is a sestic pure scalar potential, whereas $S_{\mathrm{pot}}^{\mathrm{ferm}}$ contains quartic couplings between scalars and fermions. The interested reader can find their expressions for instance in \cite{Benna:2008zy}. 

The theory exhibits extended ${\cal N}=6$ supersymmetry, with $SU(4)$ being the corresponding R-symmetry group. It can be studied perturbatively in the coupling constant $\l = N/k$ for $N \ll k$. In the opposite regime, for $N \gg k^5$ the model is dual to M--theory on ${\rm AdS}_4 \times S^7/Z_k$, whereas in the range $k \ll N \ll k^5$  it corresponds to Type IIA on ${\rm AdS}_4 \times CP^3$. 

As already mentioned in the introduction, in the ABJM theory it is possible to define two different kinds of Wilson operators. The first is the set of bosonic WLs which correspond to generalized connections that include couplings to the $(C_I, \bar{C}^I)$ scalars \cite{Berenstein:2008dc,Drukker:2008zx,Chen:2008bp}. The second set of operators, which does not have analogue in four dimensions, is made by fermionic WLs and involve couplings also to fermions. Though the possibility of introducing couplings to fermions comes simply from dimensional considerations, it turns out to be crucial for enhancing the number of preserved supersymmetries. This was originally discussed in \cite{Drukker:2009hy} where it was shown that while a bosonic WL can be at most 1/6 BPS, the addition of fermionic couplings can increase the BPS degree to 1/2. 

The complete classification of bosonic and fermionic WLs generically featured by parametric couplings to scalar and fermions has been given in \cite{Ouyang:2015iza,Ouyang:2015bmy} (and reviewed in \cite{Mauri:2017whf}). Among them we find the latitude WLs that have been introduced in \cite{Bianchi:2014laa} (also inspired by \cite{Cardinali:2012ru}). This particular set of operators depend on a real parameter $\nu \in [0,1]$ which appears in the internal couplings with matter. As discussed in \cite{Bianchi:2014laa}, this parameter can also be ascribed to a deformation of the contour $\G$ from the maximal circle on $S^2 \subset S^3$ to a latitude circle described by coordinates
$x^\mu = (\sqrt{1 - \nu^2}, \nu \cos{\tau}, \nu \sin{\tau})$. This is the reason why we call these operators ``latitude'' WLs. 

These operators are explicitly constructed according to the following prescription. 

\vskip 5pt
\noindent
{\bf Bosonic WLs}: These are the natural generalization to three dimensions of the WL in (\ref{WL4D}). Choosing a convenient normalization, they are defined as
\bea\label{WLb}
&& W_B (\nu) =    \frac{1}{N} \, \Tr \, P\,  \exp{\left( - i \int_\G d\tau {\cal L}_B (\nu, \tau) \right)}      \qquad \quad {\cal L}_B(\nu, \tau) =  \dot{x}^\mu A_\mu - \frac{2\pi i}{k}  |\dot{x}| \, M_J^{\; I} (\nu,\tau)  \, C_I \, \bar{C}^J    \nonumber \\
&& \hat{W}_B (\nu) = \frac{1}{N} \, \Tr \, P\,  \exp{\left( - i \int_\G d\tau \hat{{\cal L}}_B (\nu, \tau) \right)}   \qquad  \quad    \hat{\cal L}_B(\nu, \tau) = \dot{x}^\mu \hat{A}_\mu  - \frac{2\pi i}{k}  |\dot{x}| \, M_J^{\; I} (\nu,\tau)  \,  \bar{C}^J \, C_I   \nonumber \\
\eea
where $\G$ is a closed contour in $S^2$ and the matrix coupling is given by
\bea
&&  \mbox{\small $\!  M_{J}^{\ I} (\nu,\tau)=\left(\!\!
\begin{array}{cccc}
 - \nu  & e^{-i \tau } 
   \sqrt{1-\nu ^2} & 0 & 0 \\
e^{i \tau }  \sqrt{1-\nu ^2}
   & \nu  & 0 & 0 \\
 0 & 0 & -1 & 0 \\
 0 & 0 & 0 & 1 \\
\end{array}\!
\right)$}
\eea
For generic values of the $\nu$ parameter these operators are 1/12 BPS, that is they preserve two independent linear combinations ${\cal Q}_1(\nu)$ and  ${\cal Q}_2(\nu)$ of the original ${\cal N}=6$ supercharges, whose coefficients depend explicitly on $\nu$ \cite{Bianchi:2014laa}. For the special value $\nu=1$ the matrix $M_J^{\ I}$ becomes $\tau$-independent and diagonal, $M_J^{\ I} = {\rm diag}(-1,1,-1,1)$. In this case the operator preserves a $SU(2) \times SU(2)$ subset of the original R-symmetry group and supersymmetry is enhanced to 1/6 BPS. In fact, these operators coincide with the bosonic 1/6 BPS WL on the maximal circle on $S^2$ introduced in \cite{Drukker:2008zx,Chen:2008bp}. 

\vskip 5pt
\noindent
{\bf Fermionic WLs}: In this case the generalized connection gets promoted to a $U(N|N)$ superconnection and the operators read
\begin{eqnarray}\label{WLf}
&& \qquad W_F(\nu) =  {\cal R} \, \Tr \, P \, \exp{\left( - i \int_\G d\tau {\cal L}(\nu, \tau) \right) }  \nonumber \\
&& {\cal L}(\nu, \tau) \, = \, \left( \begin{array}{ccc}
\dot{x}^\mu A_\mu  - \frac{2\pi i}{k}  |\dot{x}| M_J^{\; I} (\nu,\tau)  \, C_I \bar{C}^J & -i \sqrt{\frac{2\pi}{k}} |\dot{x}| \eta_I(\nu,\tau) \, \bar{\psi}^I \\
 -i \sqrt{\frac{2\pi}{k}}  |\dot{x}| \psi_I \,  \bar{\eta}^I (\nu,\tau) & \dot{x}^\mu  \hat{A}_\mu 
- \frac{2\pi i}{k}  |\dot{x}| M_J^{\; I}(\nu,\tau)  \, \bar{C}^J C_I 
\end{array}\right) 
\end{eqnarray}
where $\G$ is a closed contour in $S^2$ and
\bea
\hspace{-0.3cm} \mbox{\small $M_{I}^{ \  J} (\nu, \tau) \!=\!\!\left(\!\!
\begin{array}{cccc}
 - \nu  & e^{-i \tau } 
   \sqrt{1-\nu ^2} & 0 & 0 \\
e^{i \tau }  \sqrt{1-\nu ^2}
   & \nu  & 0 & 0 \\
 0 & 0 & 1 & 0 \\
 0 & 0 & 0 & 1 \\
\end{array}\!
\right)$} \, , \quad \mbox{\small $\begin{array}{l}\eta_I^\alpha (\nu,\tau) = \frac{e^{\frac{i\nu \tau}{2}}}{\sqrt{2}}\left(\!\!\!\begin{array}{c}\!\sqrt{1+\nu}\\ -\sqrt{1-\nu} e^{i\tau}\\0\\0 \!\end{array}\!\!\right)_{\!\! \! I} \! \! \!\!(1, -i e^{-i \tau})^\alpha
\end{array}$}\!\!
\eea
Here we have chosen the convenient normalization factor 
\beq
{\cal R} = \frac{i}{2N \sin{(\pi \nu/2)}}
\eeq
which is meaningful as long as $\nu \neq 0$. For $\nu =0$ one can simply choose ${\cal R}= 1$.

In general, this class of operators is 1/6 BPS. They preserve four $\nu$-dependent linear combinations ${\cal Q}_{1,2,3,4}(\nu)$ of the original supersymmetry charges. Enhancement of supersymmetry occurs at $\nu=1$ where the matrix becomes constant and diagonal, $M_{I}^{ \  J} = {\rm diag}(-1,1,1,1)$ and preserves a $SU(3)$ subgroup of the R-symmetry group. In this case the operator coincides with the 1/2 BPS WL on the maximal circle introduced in \cite{Drukker:2009hy}.

Both kinds of operators have a well-defined limit for $\nu \to 0$, where they reduce to Zarembo-like Wilson loops \cite{Zarembo:2002an}. 

As discussed in \cite{Drukker:2009hy,Bianchi:2014laa}, classically the fermionic WL in (\ref{WLf}) is cohomologically equivalent to the following linear combination of bosonic latitudes
\beq\label{cohomo}
W_F(\nu) =  \frac{ e^{-\frac{i\pi  \nu}{2}} \, W_B(\nu) - e^{\frac{i \pi \nu}{2}} \, \hat W_B(\nu) }{ e^{-\frac{i\pi  \nu}{2}} - e^{\frac{i \pi \nu}{2}}} \, + \, {\cal Q}(\nu)\!-\!{\rm exact \; term}
\eeq
where ${\cal Q}(\nu)$ is a linear combination of superpoincar\'e and superconformal charges preserved by all the operators. If this equivalence survives at quantum level,  taking the vev of both sides of this identity we can determine $\langle W_{F}  (\nu) \rangle$  as a combination of the bosonic $\langle W_{B}  (\nu) \rangle$, $\langle \hat{W}_{B}  (\nu) \rangle$. 
However, in three dimensions the problem of understanding how the classical cohomological equivalence gets implemented at quantum level is strictly interconnected with the problem of understanding framing.

In three-dimensional pure Chern-Simons theories WL expectation values are affected by finite regularization ambiguities associated to singularities arising when two fields running on the same closed contour clash \cite{Witten:1988hf}. In perturbation theory, this phenomenon is ascribable to the use of point--splitting regularization to define propagators at coincident points \cite{Guadagnini:1989am, Alvarez:1991sx}. For ordinary WLs one allows one endpoint of the gluon propagator to run on the original closed path $\Gamma$, and the other to run on a framing contour $\Gamma_f$, infinitesimally displaced from $\Gamma$.
Then the one--loop Chern--Simons contribution is proportional to the Gauss linking integral 
\begin{equation}\label{eq:gauss}
\frac{1}{4\pi} \oint_{\G} dx^\m \oint_{\G_f} dy^\n \; \varepsilon_{\m\n\rho} \frac{(x-y)^\rho}{|x-y|^3} \, \equiv \, f
\end{equation}
which evaluates to an integer $f$ (the framing number). This is a topological invariant and corresponds to the winding number of the framing contour around the original one.  Going at higher orders this result exponentiates and the total effect of framing (scheme) dependence amounts to a controllable phase $e^{i \pi f \l}$. In non-topological Chern-Simons-matter theories framing effects still appear \cite{Bianchi:2016yzj, Bianchi:2018bke} but they no longer produce a phase factor, rather they give rise to phase functions $e^{i \phi(\l)}$ which can be computed order by order in perturbation theory.
 
Going back to the problem of understanding how the cohomological equivalence is implemented at quantum level, a two-loop computation done in dimensional regularization, thus corresponding to framing zero, shows that identity (\ref{cohomo}) holds identical for the vev's only if these are interpreted as expectations values at framing $\nu$ \cite{Bianchi:2014laa}. Precisely, if up to two loops we define framing-$\nu$ quantities as (the subscript indicates framing) 
\begin{align}
\label{bosonicWfram}
& \langle W_B (\nu)\rangle_\nu \equiv   e^{i \pi\nu \l}\,\langle W_B(\nu)\rangle_0  \quad , \quad 
\langle \hat{W}_B (\nu)\rangle_\nu  \equiv e^{-i\pi\nu \l}\,\langle\hat W_B(\nu)\rangle_0
\nonumber \\
& \langle W_F(\nu)\rangle_\nu \equiv \langle W_F(\nu)\rangle_0
\end{align}
then the quantum cohomological equivalence reads
\beq\label{equivalence}
\langle W_F(\nu) \rangle_\nu =    \frac{e^{-\frac{i\pi \nu}{2}}\, \langle W_B(\nu) \rangle_\nu \, - e^{\frac{i \pi  \nu}{2}} \, \langle \hat   W_B(\nu) \rangle_\nu }{ e^{-\frac{i\pi  \nu}{2}} - e^{\frac{i \pi \nu}{2}}}
\eeq
Identification (\ref{bosonicWfram}) for expectation values at framing-$\nu$ has been confirmed by a genuine three-loop calculation of $\langle W_B (\nu)\rangle$ and 
$\langle\hat W_B(\nu)\rangle$ done with point-splitting regularization at winding $\nu$ \cite{Bianchi:2018bke}. For $\nu=1$ identity (\ref{equivalence}) has been confirmed perturbatively, up to two loops \cite{Bianchi:2013zda,Bianchi:2013rma,Griguolo:2013sma}. 

These results show that framing, first discovered as a topological property of topological theories like pure Chern-Simons, survives also in non-topological theories, but there it is no longer an integer. Moreover, as already mentioned, the two-loop phase factors in (\ref{bosonicWfram}) get corrected at higher orders giving rise to non-trivial framing functions. As discussed in \cite{Bianchi:2018bke} and reviewed below, for $\nu=1$ the framing function is entirely due to framing effects. Instead, for generic values $\nu$ of the latitude the total phase contains {\em all} but {\em not only} framing ambiguities. In fact, already at three loops framing independent imaginary contributions appear, which concur in reconstructing the phase.

To conclude this section we report the most updated perturbative result for the bosonic and fermionic WLs. At finite $N$ and framing $\nu$ the bosonic operators read \cite{Bianchi:2018bke}
 \begin{align}\label{eq:Wb}
\langle W_B(\nu) \rangle_\nu &= 1+i \pi \nu \frac{N}{k}+\frac{\pi ^2}{6 k^2} \left(2N^2 +1\right) 
+\frac{i \pi ^3 N}{6 k^3} \left[N^2 \left(\nu^3 -  \nu +1 \right) + \nu^3  + 3\nu \right]+O\left(k^{-4}\right) \nonumber \\
\langle \hat W_B(\nu) \rangle_\nu &= 1-i \pi \nu \frac{N}{k}+\frac{\pi ^2}{6 k^2} \left(2N^2 +1\right) 
-\frac{i \pi ^3 N}{6 k^3} \left[N^2 \left(\nu^3 -  \nu +1 \right) + \nu^3  + 3\nu \right]+O\left(k^{-4}\right)
\end{align}
whereas, using (\ref{equivalence}) we find
\begin{align}\label{eq:Wf}
\hspace{-0.3cm} \langle W_F(\nu) \rangle_\nu  
&=   1 - \frac{ \pi  \nu  N^2 }{k}\, \cot\frac{\pi \nu}{2} - \, \frac{\pi^2 N}{6k^2} \,  \left( 2N^2+1 \right)  
 -\frac{ \pi ^3 \nu  N^2 }{6 k^3}  
 \left[\nu ^2 \left(N^2+1\right) + 3\right] \cot\frac{\pi \nu}{2} 
+O\left(k^{-4}\right)  
\end{align}
A useful {\em Mathematica} package for weak coupling expansions of generic WLs in ABJM theory can be found  in \cite{Preti:2017fjb}.

\section{Matrix Model and exact results for latitude Wilson loops}\label{MM}

The operators defined in the previous section possess a sufficient degree of supersymmetry to allow for the use of localization techniques in determining their vacuum expectation values. This program involves localizing the ABJ(M) theory on $S^3$ and has been accomplished in \cite{Kapustin:2009kz} for the 1/6 BPS WL, that is the operator in (\ref{WLb}) with $\nu=1$. The functional integral which computes its vev reduces to a Matrix Model that can be evaluated at weak coupling, at strong coupling and exactly in the large $N$ limit. 
In the next subsection we review the results of \cite{Kapustin:2009kz} for the $\nu=1$ case, while we postpone the discussion of the more general $\nu \neq 1$ case to the subsequent section.

\subsection{The $\nu = 1$ case} \label{undeformed}

For $\nu = 1$ the bosonic WL (\ref{WLb}) enhances to a 1/6 BPS operator $W_B \equiv W_B(\nu = 1)$ \cite{Drukker:2008zx,Chen:2008bp}, whereas the fermionic one in (\ref{WLf}) becomes 1/2 BPS, $W_F \equiv W_F(\nu = 1)$ \cite{Drukker:2009hy}. In this case, the classical comohological equivalence (\ref{cohomo}) reads \cite{Drukker:2009hy}
\beq\label{cohomo1}
W_F =  \frac{ W_B + \hat W_B }{2} \, + \, {\cal Q}\!-\!{\rm exact \; term}
\eeq
The Matrix Model computing the bosonic vevs has been found in \cite{Kapustin:2009kz} by localizing the path integral with ${\cal Q}$. They are given by
\beq\label{MM1-1}
\langle W_B \rangle_1 = \left\langle \frac{1}{N} \sum_{a=1}^{N} e^{2\pi\, \lambda_{a}} \right\rangle \qquad \qquad 
\langle \hat W_B \rangle_1 = \left\langle \frac{1}{N} \sum_{a=1}^{N} e^{2\pi\, \mu_{a}} \right\rangle
\eeq
where the expectation values are evaluated and normalized using the matrix model partition function
\bea \label{MM1-2}
Z = \int \prod_{a=1}^{N}d\lambda _{a} \ e^{i\pi k\lambda_{a}^{2}}\prod_{b=1}^{N}d\mu_{b} \ e^{-i\pi k\mu_{b}^{2}}   \; \; 
 \frac{\displaystyle\prod_{a<b}^{N}\sinh^2  \pi (\lambda_{a}-\lambda _{b})  \prod_{a<b}^{N}\sinh^2 \pi (\mu_{a}-\mu_{b}) }{\displaystyle\prod_{a=1}^{N}\prod_{b=1}^{N}\cosh^2 \pi (\lambda _{a}-\mu_{b})} 
\eea 
In these expressions $a,b$ indices label the $N+N$ eigenvalues $(\l_a, \mu_a)$ of $U(N) \times U(N)$ Cartan subalgebra. 

As previously discussed, when computing WLs in three dimensions an important issue to keep under control is framing. As argued in \cite{Kapustin:2009kz}, the Matrix Model always computes $\langle W_B \rangle$, $\langle \hat W_B \rangle$ at framing one\footnote{This is the meaning of the subscript ``1'' in (\ref{MM1-1}).}. This can be traced back to the fact that the only point-splitting regularization which does not break supersymmetry on $S^3$ corresponds to taking the original path and the framed one to belong to a Hopf fibration of the sphere.  

Since the Matrix Model is explicitly invariant under the action of ${\cal Q}$,  taking the expectation value of the comohological indentity in (\ref{cohomo1}) we 
immediately obtain the quantum result for the fermionic WL, at framing one and as a function of the results (\ref{MM1-1})
\beq
 \langle W_F \rangle_1 =    \frac{\langle W_B \rangle_1 + \langle \hat   W_B \rangle_1 }{2}
\eeq
The expressions for $\langle W_B \rangle$ and $\langle \hat W_B \rangle$ are in general complex functions of the $\l= N/k$ coupling, and can be written as 
\beq\label{complex}
 \langle W_B \rangle_1 = e^{i \Phi_B(\l)} |\langle W_B(\l) \rangle| \qquad \qquad  \langle \hat{W}_B \rangle_1 = e^{-i \Phi_B(\l)} |\langle \hat{W}_B(\l) \rangle|
\eeq
where $ |\langle W_B (\l) \rangle|, |\langle \hat{W}_B(\l) \rangle|$ are framing independent, whereas all the framing effects are encoded in the framing function $\Phi_B(\l)$. This can be established by expanding these expressions at small $\l$ \cite{Kapustin:2009kz,Marino:2009jd,Drukker:2010nc} and comparing them with a genuine perturbative calculation done at framing $f =1$. This has been done up to three loops in  \cite{Bianchi:2016yzj,Bianchi:2018bke}. In particular, $\Phi_B(\l)$ turns out to be an odd expansion, whose first few terms read
\beq\label{framing}
\Phi_B = \pi \l - \frac{\pi^3}{2} \l^3  + {\cal O} \left( \l^5 \right)
\eeq
We note that the lowest order correction coincides with the phase introduced in definitions (\ref{bosonicWfram}) when we set $\nu=1$. 

Since in this case all the framing effects are enclosed into the phase function $\Phi_B$, the moduli $ |\langle W_B \rangle|, |\langle \hat{W}_B \rangle|$ correspond to the expectation values computed at framing zero, that is in ordinary perturbation theory with a regularization prescription alternative to point-splitting. Using dimensional regularization, these expressions have been computed up to order $\l^2$ \cite{Bianchi:2014laa}, and their contributions at cubic order can be inferred from the results in \cite{Bianchi:2016yzj,Bianchi:2018bke} by setting $f=0$ there.

\subsection{The $\nu \neq 1$ case} 

The generic parametric WLs (\ref{WLb}, \ref{WLf}) deserve a separate discussion. In fact, in this case the lower degree of supersymmetry preserved by these operators leads to kind of different features in the general structure of their vevs. We are primarily interested in finding a Matrix Model prescription generalizing the previous one.  

As happens in the $\nu=1$ case, if we were able to localize the path integral for $\langle W_B(\nu) \rangle, \langle \hat{W}_B(\nu) \rangle$ using the ${\cal Q}(\nu)$ supercharge that appears in the cohomological identity (\ref{cohomo}), taking the vev of that identity we would immediately obtain $\langle W_F(\nu) \rangle$ as a function of the bosonic vevs. However, up to date, the localization procedure for this general case has not been done yet. In particular, the main difficulty stems from the fact that ${\cal Q}(\nu)$ is not chiral as it is in the $\nu=1$ case and therefore the procedure of \cite{Kapustin:2009kz} is not easily generalizable.  

In the absence of a known localization procedure compatible with (\ref{cohomo}), the perturbative results (\ref{bosonicWfram}--\ref{eq:Wf}) can be somehow inspiring for guessing the structure of the final Matrix Model which should arise from localization. Using these results as root guidance, together with the fact that the Matrix Model should reduce to (\ref{MM1-1}, \ref{MM1-2}) for $\nu=1$, the following Matrix Model prescription has been proposed in \cite{Bianchi:2018bke} for computing the bosonic latitude WLs at framing $\nu$
\beq\label{conjecture}
\langle W_B(\nu) \rangle_{\nu}  = \left\langle \frac{1}{N} \sum_{a=1}^{N} e^{2\pi\,  \sqrt{\nu} \, \lambda_{a}} \right\rangle \qquad \qquad
\langle \hat{W}_B(\nu) \rangle_{\nu}  = \left\langle \frac{1}{N} \sum_{a=1}^{N} e^{2\pi\,  \sqrt{\nu} \, \mu_{a}} \right\rangle
\eeq
where now the average is evaluated and normalized with the following partition function
\bea\label{zeta}
&& Z (\nu)= \int \prod_{a=1}^N d\lambda _a \ e^{i\pi k\lambda_a^2}\prod_{b=1}^N d\mu_b \ e^{-i\pi k\mu_b^2}   \\
&& \times  \, 
 \frac{\displaystyle\prod_{a<b}^N \sinh \sqrt{\nu} \pi (\lambda_a -\lambda _b) \sinh \frac{\pi (\lambda_a-\lambda _b)}{ \sqrt{\nu}} \prod_{a<b}^{N} \sinh \sqrt{\nu} \pi (\mu_a-\mu_b) \sinh \frac{\pi (\mu_a-\mu_b)}{ \sqrt{\nu} } }{\displaystyle\prod_{a=1}^N \prod_{b=1}^N \cosh  \sqrt{\nu} \pi (\lambda _a -\mu_b) \cosh \frac{\pi (\lambda _a-\mu_b)}{ \sqrt{\nu} } }\nonumber
\eea
As before, the integral is over a set of $(\l_a, \mu_a)$  eigenvalues of the Cartan matrices of $U(N) \times U(N)$. 

In addition, relying on the quantum cohomological equivalence, the prescription to determine $\langle W_F(\nu) \rangle$ reads
\beq \label{cohomo2}
\langle W_F(\nu) \rangle_\nu =  \frac{ e^{-\frac{i\pi  \nu}{2}} \, \langle W_B(\nu) \rangle_\nu - e^{\frac{i \pi \nu}{2}} \, \langle  \hat W_B(\nu) \rangle_\nu }{ e^{-\frac{i\pi  \nu}{2}} - e^{\frac{i \pi \nu}{2}}}  
\eeq
The Matrix Model provides results for the bosonic latitudes that are in general complex functions of the coupling $\l$. They can then be expressed as
\beq\label{complex2}
\langle W_B(\nu) \rangle_\nu = e^{i \, \Phi_B(\nu)} |\langle W_B(\nu) \rangle_\nu| \quad , \quad 
\langle \hat W_B(\nu) \rangle_\nu = e^{-i \, \Phi_B(\nu)} |\langle \hat W_B(\nu) \rangle_\nu| 
\eeq
whereas $\langle W_F(\nu) \rangle_\nu$ computed from (\ref{cohomo2}) is a real quantity. It is important to observe that for $\nu \neq 1$ the phase $\Phi_B(\nu)$ includes all, but {\em not only} framing effects. In fact, a perturbative evaluation of $\langle W_B(\nu) \rangle_\nu$ at framing $f$ shows that already at three loops an imaginary contribution arises, which is framing independent and proportional to $(\nu^2 - 1)$ \cite{Bianchi:2018bke}. Remarkably, this contribution is strictly related to an anomalous behaviour of two-point correlation functions in the defect CFT defined on the bosonic Wilson line. 

These expressions have been computed at large $N$ in the strong coupling limit, using the Fermi gas approach.  Applying a genus expansion in powers of the string coupling $g_s = \tfrac{2\pi i}{k}$, and introducing the new variable $\kappa$ through the identity
\begin{equation}
\l = \frac{\log^2 \kappa}{2\pi^2} + \frac{1}{24} + O\left(\kappa^{-2}\right)
\end{equation}
the genus-zero terms (leading order in $1/k$) read \cite{Bianchi:2018bke}
\begin{equation} \label{eq:genus0}
\langle W_B(\nu) \rangle_\nu\,\big|_{g=0} = \frac{-\kappa ^{\nu }\, \Gamma \left(\frac{\nu -1}{2}\right) \Gamma \left(\frac{\nu +1}{2}\right)+ i\, \pi\,  \kappa  \left(1+i\, \tan \frac{\pi  \nu }{2}\right) \Gamma (\nu +1)}{4 \pi\, \Gamma (\nu +1)}
\end{equation}
with $\langle  \hat W_B(\nu) \rangle$ given simply by the hermitian conjugate of $\langle W_B(\nu) \rangle$, and 
\begin{equation}\label{eq:fermionicgenus0}
\langle W_F(\nu) \rangle_\nu\,\big|_{g=0} = -i\, \frac{2^{-\nu -2}\, \kappa ^{\nu }\, \Gamma \left(-\frac{\nu }{2}\right)}{\sqrt{\pi}\, \Gamma \left(\frac{3}{2}-\frac{\nu }{2}\right)}
\end{equation}
 
Proposal (\ref{conjecture}) for the latitude Matrix Model relies on a number of  strong consistency checks. First of all, it is the simplest deformation of (\ref{MM1-1}) that reduces to that expression for $\nu=1$.

 A first non-trivial check concerns the partition function (\ref{zeta}) itself. If the interpretation of this Matrix Model as the result of localizing with the $\nu$-dependent supercharge appearing in (\ref{cohomo}) is correct, then expression (\ref{zeta}) should provide the ordinary $\nu$-independent partition function of the ABJM model. In fact, ${\cal Q}(\nu)$ is a supersymmetry of $Z$ for any $\nu$ and the result for the partition function should not depend on the localizing supercharge that we use. This has been successfully checked in \cite{Bianchi:2018bke} where it has been shown that expression
(\ref{zeta}) can be rearranged in such a way that the $\nu$ dependence disappears completely and it ends up coinciding with the ABJM partition function on $S^3$ \cite{Kapustin:2009kz}. Of course, such manipulations no longer work when we insert the WL exponential.

Important checks come from comparing the Matrix Model results at weak and strong couplings with alternative calculations. At weak coupling, its expansion perfectly matches perturbative results (\ref{eq:Wb}, \ref{eq:Wf}) done at framing $\nu$. This confirms the intuition that localization should compute WLs at non-integer framing $\nu$. At strong coupling, the leading exponential behavior of (\ref{eq:fermionicgenus0})   
\beq
\langle W_F(\nu) \rangle \sim e^{\pi \nu \sqrt{2\l}}
\eeq
reproduces the holographic prediction found in \cite{Correa:2014aga}.
Moreover, in \cite{Medina-Rincon:2019bcc}  the ratio $\frac{\langle W_F(1) \rangle}{ \langle W_F(\nu) \rangle}\Big|_{g=0}$ has been computed holographically at strong coupling, at the next-to-leading order, and the result perfectly matches the Matrix Model prediction from (\ref{eq:fermionicgenus0}).

Finally, due to the particular $\nu$ dependence of $Z$ in (\ref{zeta}), the expectation values (\ref{conjecture}) satisfy the functional identity
\beq\label{eq:identity}
\partial_\nu \log{\left( \langle W_B(\nu) \rangle_\nu + \langle \hat W_B(\nu) \rangle_\nu \right)} = 0  
\eeq
In other words, the real part of the average $\langle W_B(\nu) \rangle_\nu$ is independent of $\nu$. This peculiar property is going to be useful for the discussion in section \ref{Bframing}.

\section{An exact prescription for the Bremsstrahlung functions}\label{B}

As reviewed above, both in four and three-dimensional superconformal models localization provides exact prescriptions for computing circular WL averages exactly. It is then tempting to exploit circular WLs for determining other physical observables of interest in these theories. A remarkable example of observable is the Bremsstrahlung function, which has been shown to be strictly related to circular WLs through non-trivial identities that rely on the power of (super)conformal symmetry. 
After briefly reviewing what happens in the ${\cal N}=4$ SYM case, we focus on the configuration of Bremsstrahlung functions in ABJM theory and their relation with latitutude WLs. 

\begin{figure}
     \hspace{4.5cm} \includegraphics[width=.4\textwidth]{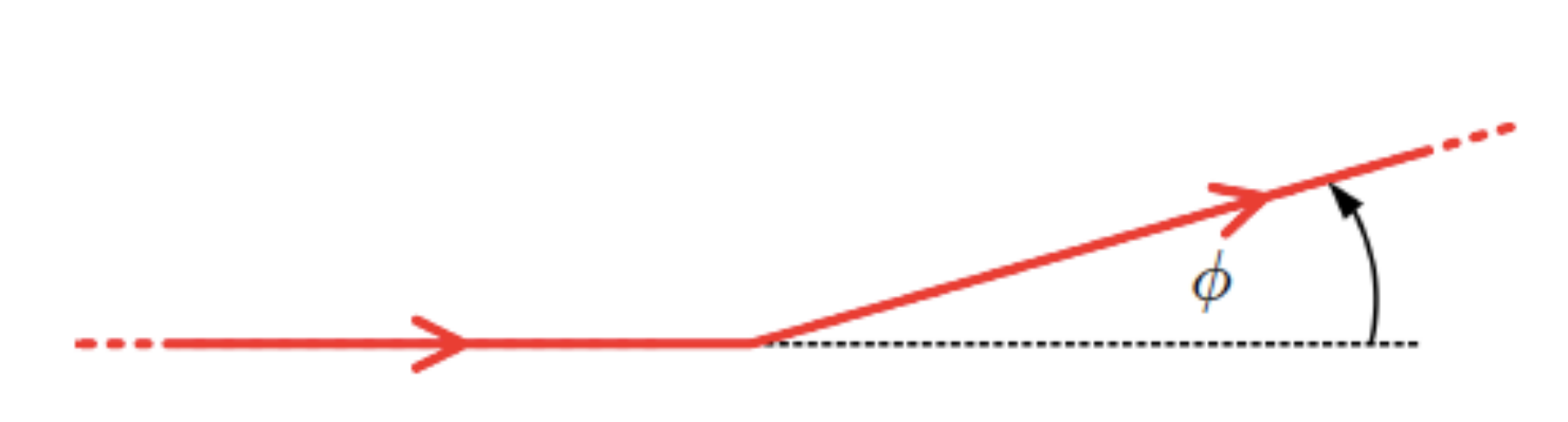}
     \caption{A cusped contour for a Wilson operator.}
     \label{fig1}
     \end{figure}

Generalising the well-known Larmor-Li\'enard  formula of QED, in a generic gauge theory the Bremsstrahlung function $B$ defines the energy lost by a massive quark slowly moving in a gauge background with velocity $\omega$,
\beq\label{B}
\Delta E = 2\pi \, B \int dt \,  \dot{\omega}^2 \, ,\qquad \qquad |\omega| \ll 1
 \eeq
In CFT it is also related to the cusp anomalous dimension $\G_{cusp}$ which weights the singular part of a Wilson operator at a cusp (see figure \ref{fig1}). In fact, close to the cusp short distance singularities appear, which exponentiate as
\beq
\langle W^\angle \rangle_{\phi} \sim e^{-\G_{cusp} (\phi) \log{\frac{L}{\e}} }  \
\eeq
where $L$ is a large distance regulator, while $\e$ is the UV regulator. For small angles, $\phi \ll 1$,  the cusp anomalous dimension behaves as 
$ \G_{cusp} (\phi) \sim  - B \, \phi^2$ \cite{Correa:2012at}, where $B$ is the Bremsstrahlung function defined in (\ref{B}). In general it is a non-trivial function of the coupling constant of the theory. 

This quantity has been first defined and computed in the ${\cal N}=4$, $SU(N)$ SYM theory for the generalized Wilson-Maldacena operator (\ref{WL4D}). 
In this case, since the operator is featured by both a geometric cusp angle $\phi$ as in figure \ref{fig1} and an internal angle $\th$ which drives the coupling to the scalar fields, for the singular part of its vev we have
\beq
\langle W^\angle(\th) \rangle_{\phi} \sim e^{-\G_{cusp} (\phi, \th) \log{\frac{L}{\e}} }  \
\eeq
and for small angles one can prove that  \cite{Correa:2012at}
\beq
\G_{cusp} (\phi,\th) \underset{\theta, \phi \ll 1}{\sim}  B \, (\th^2 - \phi^2) 
\eeq
In particular, this expression vanishes for  $\th^2 = \phi^2$ where the divergences disappear and the cusp becomes BPS.

Although in principle $B$ could be computed directly from the cusp anomalous dimension, this is in general obstructed by the fact that  the perturbative evaluation of $\G_{cusp}$ is not an easy task already at low orders. However, exploiting the line-to-circle mapping in CFTs, an exact prescription has been proposed \cite{Correa:2012at} for computing this quantity in terms of the 1/2 BPS circular Wilson loop
\beq\label{4DB} 
B = \frac{1}{2\pi^2}\, \l \partial_{\l} \log  \langle W_{1/2} \rangle   \; , \qquad \quad \l = g_{YM}^2 N
\eeq
where $ W_{1/2}$ is the operator in (\ref{WL4D}) evaluated on the maximal circle in $S^2$, for which localization predicts the exact expression (\ref{4Dexact}). Therefore, applying identity (\ref{4DB}) to this result leads to an exact expression for $B$. This result has been checked at weak coupling against a genuine perturbative calculation of $\G_{cusp}$ up to four loops \cite{Correa:2012at,Correa:2012nk,Henn:2013wfa} and at strong coupling up to one loop \cite{Drukker:2011za,Forini:2010ek}. 
Remarkably, it has been also obtained solving a TBA system of integral equations in a suitable limit \cite{Drukker:2012de}-\cite{Gromov:2015dfa}. Therefore, the observable $B$ plays also a crucial role in testing integrability underlying the AdS/CFT correspondence. 

In ABJM theory, since there are bosonic and fermionic WLs we can define different types of cusped operators and consequently different types of Bremsstrahlung functions \cite{Griguolo:2012iq,Lewkowycz:2013laa}. 

Computing the divergent contributions to a fermionic, 1/2 BPS operator close to a cusp (see figure \ref{fig1}) we obtain
\beq
\langle W_F^\angle (\theta) \rangle_{\phi} \sim e^{-\G^{1/2}_{cusp} (\phi, \theta) \log{\frac{L}{\e}} } \; , \qquad  \qquad  \G^{1/2}_{cusp} (\phi,\theta) \underset{\phi, \theta \ll 1}{\sim} B_{1/2} (\theta^2 -\phi^2)
\eeq
where $\theta$ is an internal angle that describes possible relative rotations of the matter couplings between the Wilson loops defined on the two semi–infinite lines. It is important to stress that $B_{1/2}$ appears as a common factor due to the fact that for $\theta^2 = \phi^2$ the cusp anomalous dimension vanishes and the cusped operator becomes BPS. 

Instead, if we study the short distance behaviour of a 1/6 BPS bosonic operator near a cusp, we find that no BPS condition enhances in this case and we are forced to define two different $B$ functions
\beq
\langle W_B^\angle (\theta) \rangle_{\phi} \sim e^{-\G^{1/6}_{cusp} (\phi, \theta) \log{\frac{L}{\e}} } \; ,  \qquad \qquad  \G^{1/6}_{cusp} (\phi, \theta) \underset{\phi, \theta \ll 1}{\sim}  B_{1/6}^{\theta }\, \theta^2  -  B_{1/6}^{\phi }\, \phi^2 
\eeq
All the $B$'s are in general functions of the coupling constant $\l$ and require specific determination. To this end, as in the four-dimensional case, it is crucial to relate these quantities to observables whose vevs are known exactly from localization. 
  
This problem has been originally addressed in \cite{Lewkowycz:2013laa}, where the following prescription for computing $B_{1/6}^{\phi}$ in terms a $m$-winding circular 1/6 BPS bosonic WL was proposed
\beq\label{m}
B_{1/6}^{\phi} = \frac{1}{4\pi^2}\, \partial_m \log \left| \, \langle W_B^{m} \rangle\, \right|\,\, \Big|_{m=1} 
\eeq
A similar prescription has been later proposed for computing $B_{1/2}$ \cite{Bianchi:2014laa} and $B_{1/6}^{\theta}$ \cite{Correa:2014aga} in terms of fermionic (\ref{WLf}) and bosonic (\ref{WLb}) latitude WLs, respectively
\beq\label{prescription}
B_{1/2} = \frac{1}{4\pi^2}\, \partial_{\nu} \log \left|\, \langle W_F(\nu) \rangle \, \right|\,\, \Big|_{\nu=1}     \qquad , \qquad 
B_{1/6}^{\theta} = \frac{1}{4\pi^2}\, \partial_{\nu} \log \left|\, \langle W_B(\nu) \rangle \, \right|\,\, \Big|_{\nu=1}
\eeq 
It is important to note that all these identities require taking the modulo of the BPS WL, in contrast with the analogous prescription (\ref{4DB}) in ${\cal N}=4$ SYM. This is due to the fact that in three dimensions the WL vevs acquire imaginary contributions both from framing and non-framing effects. The modulo is then necessary in order to obtain a real expression for the $B$ functions. A detailed discussion about this point can be found in \cite{Bianchi:2018bke}. 

Formulae (\ref{prescription}) have been proved in \cite{Bianchi:2017ozk} and \cite{Correa:2014aga}, respectively, using their relation with correlation functions in 
one-dimensional defect CFTs defined on the Wilson lines. Moreover, the interesting relation 
\beq
B_{1/6}^\theta = \frac12 B_{1/6}^\phi
\eeq
has been guessed in \cite{Bianchi:2017afp, Bianchi:2017ujp} from a four-loop calculation and finally proved in \cite{Bianchi:2018scb}. In particular, using equations (\ref{m}) and (\ref{prescription}), this identity implies a non-trivial relation between the $\nu$-derivative of the latitude $W_B(\nu)$ and the $m$-derivative of the $m$-winding WL. 

Perturbative checks up to two loops for $B^{\phi}_{1/6}$ and $B_{1/2}$ can be found in \cite{Griguolo:2012iq,Bianchi:2014laa}, whereas a similar check for $B^{\theta}_{1/6}$ is given in \cite{Bianchi:2014laa}. A three-loop calculation of $\G_{cusp}$ \cite{Bianchi:2017svd} provides a non-trivial check for $B_{1/2}$ at this order.
 At strong coupling, $B_{1/2}$ matches the string prediction at next-to-leading order \cite{Forini:2012bb, Aguilera-Damia:2014bqa}.

\section{Bremsstrahlung and framing}\label{Bframing}

Aa a remarkable consequence of prescriptions (\ref{prescription}) for computing the Bremsstrahlung functions a new physical interpretation of framing in 
three-dimensional Chern-Simons-matter theories arises \cite{Bianchi:2014laa,Bianchi:2017svd,Bianchi:2018scb}.

In fact, we can elaborate on the first equation in (\ref{prescription}) by substituting $\langle W_F(\nu) \rangle$ with its expression (\ref{cohomo2}) coming from the cohomological equivalence. As consequence, the fermionic $B_{1/2}$ turns out to be expressed in terms of the bosonic BPS WLs, which in turn can be written as in (\ref{complex2}). 
Now, assuming identity (\ref{eq:identity}) to be true we eventually find
\beq
B_{1/2} =  -\frac{i}{8\pi}\, \frac{\langle W_B\rangle-\langle \hat W_B\rangle}{\langle W_B\rangle+\langle \hat W_B\rangle} = \frac{1}{8\pi}\tan\Phi_B  
\eeq
where $W_B, \hat W_B$ are the underformed bosonic 1/6 BPS WLs discussed in section \ref{undeformed} and $\Phi_B$ the corresponding framing function defined in eq. (\ref{complex}). As already discussed there, for $\nu=1$ this phase contains {\em all and only} framing contributions. Therefore, we reach the conclusion that framing effects, which in topological Chern-Simons theories correspond to integer topological invariants and represent a controllable regularization scheme dependence, in non-topological Chern-Simons-matter theories are no longer numbers but functions and acquire a new physical interpretation as sources for the Bremsstrahlung function.

Similarly, elaborating the second identity in (\ref{prescription}) we easily obtain
\beq\label{eq:Bframing}
B_{1/6}^{\th} = \frac{1}{4\pi^2} \, \tan{ \Phi_B(\nu)} \; \partial_\nu \Phi_B(\nu) \Big|_{\nu=1}  
\eeq
where now $\Phi_B(\nu)$ is the generic bosonic phase function at latitude $\nu$ defined in (\ref{complex2}). In this case, as already mentioned, it contains {\em all but not only} framing contributions. This identity has been exploited to perform non-trivial checks of the whole construction. In fact, the four-loop calculation of \cite{Bianchi:2017afp,Bianchi:2017ujp} for 
$\G_{cusp}^{1/6}$ allows to determine $B_{1/6}^{\th}$ up to this order. Using equation (\ref{eq:Bframing}) this in turn provides a prediction for the expansion of 
$\Phi_B(\nu)$ up to $\l^3$  \cite{Bianchi:2018scb} (we recall that the phase function has an odd expansion in $\l$). Merging this result with the two-loop calculation of 
$|\langle W_B(\nu) \rangle|$ \cite{Bianchi:2014laa} one obtains a three-loop expansion for $\langle W_B(\nu) \rangle_\nu$. This prediction has been checked by a genuine three-loop calculation of 
$\langle W_B(\nu) \rangle$ done at framing $\nu$ \cite{Bianchi:2018bke}, and is marvellously reproduced by the Matrix Model average (\ref{MM1-1}) expanded at weak coupling.

\section{Conclusions and Perspectives}\label{conclusions}

I have reviewed recent progress in the study of generalized (latitude) bosonic and fermionic BPS Wilson operators in three-dimensional ${\cal N}=6$ ABJM theory.

The first result concerns the proposal for a $\nu$-latitude Matrix Model that computes bosonic WL averages exactly at framing $\nu$. Assuming cohomological equivalence to hold at quantum level at framing $\nu$, one can then find the result for the fermionic operator in terms of the bosonic ones. These are new exact, interpolating functions that allows to test AdS$_4$/CFT$_3$ in the large $N$ limit. In fact, for the fermionic operator the Matrix Model result expanded at strong coupling matches the holographic calculation. For the bosonic latitude no precise dual string configuration has been determined yet, though much progress has been recently done in \cite{Correa:2019rdk},  and therefore our findings constitute a brand new prediction, begging for a string theory confirmation.

Our latitude Matrix Model is expected to be the result of localizating the original path integral with a supercharge compatible with the cohomological equivalence, possibly the ${\cal Q}(\nu)$ charge itself (see eq. (\ref{cohomo})). It is then demanding to develop such a localization procedure in order to have confirmation of our proposal from first principles. However, it cannot be a straightforward generalization of the procedure developed in \cite{Kapustin:2009kz} for the $\nu=1$ case, since in this case ${\cal Q}(\nu)$ is not chiral. 

We have discussed latitude WLs in ABJM theory. However, definitions (\ref{WLb}, \ref{WLf}) can be easily generalized to the case of the $U(N_1)_k \times U(N_2)_{-k}$ Chern-Simons-matter theory (ABJ theory) \cite{ABJ}. The WL expressions are formally the same except for the overall normalizing factors that will be functions of $N_1$ and $N_2$. 
A more general Matrix Model has been proposed also for this theory \cite{Bianchi:2018bke}, which reduces to (\ref{MM1-1},\ref{MM1-2}) for $N_1=N_2$. In this case, computing the partition function, a non–trivial $\nu$-dependence survives in its phase. Its appearance could be ascribed to a Chern–Simons framing anomaly discussed in \cite{Closset:2012vg,Closset:2012vp} and leads to the conclusion that the deformation affects the partition function only in its somewhat unphysical part, whereas its modulus is $\nu$ independent. However, this finding is still not totally clear and deserves further investigation. 

We have provided an exact prescription for computing the Bremsstrahlung functions in terms of latitude WLs. These functions could be alternatively computed by exploiting the exact solvability of the model \cite{Minahan:2008hf,Gaiotto:2008cg,Gromov:2008bz,Bombardelli:2017vhk}. This would require solving a system of TBA equations, as done in ${\cal N}=4$ SYM \cite{Drukker:2012de,Correa:2012hh}. Matching localization and integrability results would be crucial for an exact check of the conjecture in \cite{Gromov:2014eha} for the interpolating function $h(\lambda)$ of the ABJM theory. Some preliminary steps in this direction involve the exact evaluation of the fermionic cusp anomalous dimension in a suitable scaling limit \cite{Bonini:2016fnc}. At strong coupling, $h(\l)$ has been tested up to two loops in the string sigma model \cite{Bianchi:2014ada}.

More generally, our Matrix Model results could be exploited for studying correlation functions of local operators in the one-dimensional defect superconformal field theory defined on the Wilson contour. For example,  derivatives of the bosonic latitude WL with respect to the parameter, $\partial^n_\nu\, \log \langle W_B(\nu) \rangle \,\Big|_{\nu=1} $, give rise to integrated correlation functions of local bilinear operators of the form $m^I_J(\tau) C_I(\tau) \bar{C}^J(\tau)$.
Knowing the explicit expression of these derivatives from the Matrix Model  and comparing them with a bootstrap evaluation of the correlation functions would provide information on the OPE data of the one-dimensional theory.  Further important questions that would be interesting to address in this context are: What is the relation between the two one-dimensional theories defined on $W_B(\nu)$ and $W_F(\nu)$? More generally, what are the implications of cohomological equivalence on the one-dimensional theories? What is the meaning of framing in the one-dimensional theories? We plan to address these questions in a near future.

\acknowledgments
First of all, I would like to thank the organizers of CORFU2019, in particular Patrizia Vitale and George Zoupanos, for putting together such an exciting Conference. Thanks to the organizers and all the participants for the relaxed and stimulating atmosphere and for the beautiful time we spent together in Corf\`u.
I would like to thank my collaborators, Marco Bianchi, Luca Griguolo, Matias Leoni, Andrea Mauri, Michelangelo Preti and Domenico Seminara, who have  crucially contributed to the accomplishment of these results. This work has been partially supported by Italian Ministero dell'Universit\`a e della Ricerca (MIUR), and by Istituto Nazionale di Fisica Nucleare (INFN) through the ``Gauge theories, Strings, Supergravity'' (GSS) research project.

\end{document}